\documentclass[12pt]{article}
\usepackage{amsmath}
\usepackage{amsthm}
\usepackage{amsfonts}
\usepackage[dvips]{epsfig}
\usepackage{graphicx}
\usepackage{amssymb}
\usepackage{cancel}
\usepackage{pstricks} 
\usepackage{lscape}
\usepackage{enumitem}
\usepackage{subcaption}

\usepackage[sort&compress,numbers, merge]{natbib}

\DeclareFontFamily{U}{mathx}{\hyphenchar\font45}
\DeclareFontShape{U}{mathx}{m}{n}{<-> mathx10}{}
\DeclareSymbolFont{mathx}{U}{mathx}{m}{n}

\newcommand{\beq}{\begin{equation}}
\newcommand{\eeq}{\end{equation}}
\newcommand{\bea}{\begin{eqnarray}}
\newcommand{\eea}{\end{eqnarray}}

\makeatletter
\newlength{\apb@width}
\newcommand{\autoparbox}[2][c]{\settowidth{\apb@width}{#2}\parbox[#1]{\apb@width}{#2}}

\newcommand{\Cen}[2]{%
  \ifmeasuring@
    #2%
  \else
    \makebox[\ifcase\expandafter #1\maxcolumn@widths\fi]{$\displaystyle#2$}%
  \fi
}
\makeatother

\definecolor{Orange}{cmyk}{0,0.61,0.87,0}
\definecolor{JungleGreen}{cmyk}{0.99,0,0.52,0}
\definecolor{OliveGreen}{cmyk}{0.64,0,0.95,0.40}
\definecolor{Brown}{cmyk}{0,0.81,1,0.60}
\definecolor{RoyalBlue}{cmyk}{0.71,0.53,0,0.12}

\allowdisplaybreaks[1]

\usepackage[colorlinks=true, linkcolor=OliveGreen, citecolor=RoyalBlue,
urlcolor=RoyalBlue]{hyperref}

\begin{document}

\vspace{-0.2in}
\begin{flushright}
{\tt KCL-PH-TH/2021-46}, {\tt CERN-TH-2021-099}  \\
{\tt ACT-1-21, MI-HET-751} \\
{\tt UMN-TH-4018/21, FTPI-MINN-21/11}
\end{flushright}

\vspace{0.5cm}
\begin{center}
{\bf \LARGE Flipped $\mathbf{g_\mu - 2}$}
\end{center}
\vspace{0.75cm}

\begin{center}{
{\bf John~Ellis}$^{a}$,
{\bf Jason~L.~Evans}$^{b}$,
{\bf Natsumi Nagata}$^{c}$, \\[0.1cm]
{\bf Dimitri~V.~Nanopoulos}$^{d}$ and
{\bf Keith~A.~Olive}$^{e}$
}
\end{center}

\begin{center}
{\em $^a$Theoretical Particle Physics and Cosmology Group, Department of
  Physics, King's~College~London, London WC2R 2LS, United Kingdom;\\
Theoretical Physics Department, CERN, CH-1211 Geneva 23,
  Switzerland;\\
National Institute of Chemical Physics and Biophysics, R\"{a}vala 10, 10143 Tallinn, Estonia}\\[0.2cm]
  {\em $^b$Tsung-Dao Lee Institute, Shanghai Jiao Tong University, Shanghai 200240, China}\\[0.2cm] 
  {\em $^c$Department of Physics, University of Tokyo, Bunkyo-ku, Tokyo
 113--0033, Japan}\\[0.2cm] 
{\em $^d$George P. and Cynthia W. Mitchell Institute for Fundamental
 Physics and Astronomy, Texas A\&M University, College Station, TX
 77843, USA;\\ 
 Astroparticle Physics Group, Houston Advanced Research Center (HARC),
 \\ Mitchell Campus, Woodlands, TX 77381, USA;\\ 
Academy of Athens, Division of Natural Sciences,
Athens 10679, Greece}\\[0.2cm]
{\em $^e$William I. Fine Theoretical Physics Institute, School of
 Physics and Astronomy, University of Minnesota, Minneapolis, MN 55455,
 USA}
 
 \end{center}

\vspace{1cm}
\centerline{\bf ABSTRACT}
\vspace{0.2cm}

{\small 
We analyze the possible magnitude of the supersymmetric contribution to $g_\mu - 2$ in a
flipped SU(5) GUT model. Unlike other GUT models which are severely constrained 
by universality relations, in flipped SU(5) the U(1) gaugino mass and the
soft supersymmetry-breaking masses of right-handed sleptons are unrelated to
the other gaugino, slepton and squark masses. Consequently, the lightest
neutralino and the right-handed smuon may be light
enough to mitigate the discrepancy between the experimental measurement of 
$g_\mu - 2$ and the Standard Model calculation, in which case they may be
detectable at the LHC and/or a 250~GeV $e^+ e^-$ collider, whereas the other
gauginos and sfermions are heavy enough to escape detection at the LHC.
}

\vspace{0.5in}

\begin{flushleft}
July 2021
\end{flushleft}
\medskip
\noindent

\newpage

\section{Introduction}

It is now 20 years since the first emergence of the discrepancy between the experimental 
value of $g_\mu - 2$ and the value calculated in the Standard Model~\cite{BNL1}. The significance of
this discrepancy has increased subsequently, with improved accuracy in the BNL
measurements~\cite{BNL2} and now the measurement by the Fermilab experiment~\cite{FNAL}, and the increased
precision in the Standard Model calculation made possible, in particular, by
improved determinations of the hadronic vacuum polarization and light-by-light
contributions~\cite{Theory}. As soon as
the first BNL result was announced, supersymmetric models were immediately proposed to explain the discrepancy~\cite{ENO,g-2}. However, the popularity of the supersymmetric explanation
has waned over the years, with the continuing lack of direct experimental evidence 
for supersymmetry, particularly at the LHC~\cite{LHCSUSY}.

However, this dampening of supersymmetric
enthusiasm is not entirely warranted. The absence at the LHC so far of squarks and 
gluinos does not bear directly on the possible masses of smuons and their sneutrino,
the lighter chargino and the
lightest neutralino, which would likely give the largest supersymmetric
contributions to $g_\mu - 2$. However, in models that 
postulate universality relations at a high grand unification (GUT)
scale, there are relations between the different gaugino
masses and between the various soft supersymmetry-breaking sfermion masses. For example, in the constrained minimal supersymmetric Standard Model (CMSSM) \cite{cmssm0}, a universal gaugino mass, $m_{1/2}$, a scalar mass, $m_0$, and a trilinear term, $A_0$, are all defined at the GUT scale and, together with the ratio of Higgs vacuum expectation values (vev), $\tan \beta$, and the sign of the $\mu$-term, define the sparticle spectrum
at the weak scale when run down from the GUT scale. Prior to the LHC searches and the discovery of the Higgs boson, the CMSSM could easily account for the $g_\mu - 2$ discrepancy \cite{ENO,g-2},
but the current experimental constraints  exclude a significant supersymmetric contribution to $g_\mu - 2$
in this and similar models~\cite{CMSSM, otherCMSSM}. 
However, if one treats the soft supersymmetry-breaking parameters as
phenomenological quantities unconstrained by GUT-scale relations, the absence of
sparticles at the LHC can be reconciled with a supersymmetric explanation of the
$g_\mu - 2$ discrepancy~\cite{pMSSM11, Sven}.~\footnote{See~\cite{otherSUSY,Cox:2021gqq} for other supersymmetric
interpretations of the $g_\mu - 2$ measurements.}

We show in this paper that a significant supersymmetric contribution
is possible in one specific GUT model, namely flipped SU(5) (FSU(5))~\cite{FSU5}.~\footnote{See~\cite{LNW}
for a previous discussion of $g_\mu - 2$ in FSU(5).} We recall that the difference
in $a_\mu \equiv (g_\mu -2)/2$ between the combination of the BNL and Fermilab
data and the data-driven value recommended in~\cite{Theory} is
$\Delta a_\mu = (251 \pm 59) \times 10^{-11}$, and that a recent lattice
calculation~\cite{Lattice} corresponds to $\Delta a_\mu = (107 \pm 69) \times 10^{-11}$.
We find a region of the FSU(5) parameter space for which the supersymmetric
contribution can reach $\Delta a_\mu|_{\rm FSU(5)} \gtrsim 140 \times 10^{-11}$, which would
reduce the discrepancy with the data-driven calculation of $a_\mu$ to below 2 standard deviations, and
remove entirely the discrepancy with the lattice calculation by the BMW
collaboration~\cite{Lattice}.

\section{Recap of the FSU(5) GUT}

Specific GUT-motivated models can interpolate between the restrictive CMSSM and the relatively unconstrained phenomenological MSSM (pMSSM) \cite{pMSSM,mc11,pMSSM11}. In a minimal SU(5) GUT,
while there is only a single universal gaugino mass, $m_{1/2} = M_{5}$, each generation of matter fields is split into ${\bf 10}$ and ${\bf \bar{5}}$ representations, which may have separate soft scalar masses, $m_{10}$ and $m_{\bar{5}}$, respectively \cite{CMSSM}. Additionally, the Standard Model Higgs fields originate from a ${\bf 5}$ and ${\bf \bar{5}}$ pair, which may also receive independent soft masses $m_{H}$ and $m_{\bar{H}}$ as in an extension of the CMSSM with non-universal Higgs masses (NUHM) \cite{nuhm2}.
The common value of the gaugino masses at the GUT scale links the electroweak gaugino masses to the gluino mass, and
the fact that both right- and left-handed (s)leptons find themselves in (super)multiplets containing (s)squarks links slepton masses to squark masses through renormalization-group running. Thus,
despite its additional degrees of freedom beyond those in the CMSSM, the SU(5) 
model 
does not resolve the $g_\mu -2$ discrepancy \cite{CMSSM}. 

On the other hand, we recall that in FSU(5) there are two independent gauge group factors:
in addition to the GUT SU(5) factor there is an `external' U(1) factor. The masses of the 
usual SU(3), SU(2) and U(1) gauginos are related by SU(5) universality 
at the GUT scale, $M_{5}$, but the mass of the `external' U(1) gaugino, $M_{X1}$, is in general 
independent. Liberated from the tyranny of GUT unification, this external
U(1) gaugino could be much lighter than the other U(1) gaugino and the Higgsinos,
enabling the lightest neutralino dark matter particle to be relatively light. 
We recall also that the right-handed sleptons are assigned
to singlet representations of FSU(5), so their soft supersymmetry-breaking
masses, $m_1$, are unrelated to those of the other sfermions, which have flipped
assignments in
$\mathbf{\overline 5}$ and $\mathbf{10}$ representations of SU(5). Therefore
the mass of the right-handed smuon, $\tilde \mu_R$, is unrelated to the masses
of the squarks and the left-handed smuon, $\tilde \mu_L$.

At one-loop order, there are contributions to $g_\mu - 2$
from a $\tilde \mu_R/\chi$ loop, a $\tilde \mu_L/\chi$ loop, and a diagram where the $\tilde \mu_R$
and $\tilde \mu_L$ mix (as well as chargino exchange diagrams). From the calculations in \cite{IN}, we find that the neutralino exchange diagrams always dominates over the chargino exchange terms, and the dominant contribution comes from
$\tilde \mu_R/\tilde \mu_L$ mixing, with the $\tilde \mu_R/\chi$ and $\tilde \mu_L/\chi$loop both sub-dominant. This is due in part to the relatively large values of $\mu$ and $A_0$ that contribute to left-right mixing. 
As we shall see, the $\tilde \mu_R$ might 
be sufficiently light, in combination with the lightest neutralino, $\chi$, to reconcile the
experimental measurement of $g_\mu - 2$ with the theoretical calculation of the
Standard Model contribution.

More specifically,
the assignments of representations and charges of each generation
of particles in the matter sector of the theory are
\begin{eqnarray}
\bar f_i (\bar{\bf 5},-3)=\left\{ U_i^c, L_i\right\} \; , \quad F_i({\bf 10},1)=\left\{ Q_i, D_i^c, N_i^c\right\} \; , \quad  l_i({\bf 1},5) =  E_i^c \; , \quad i = 1, 2, 3 \, ,
\end{eqnarray} 
where the charges are defined in the (SU(5),\ U(1)$_X$) basis. We note that there is an additional degree of freedom beyond the Standard Model contained in the ${\bf 10}$, denoted by $N^c$, which can be interpreted as a right-handed neutrino. 
In order to generate the right-handed neutrino masses, the theory contains three 
or more SU(5) singlets $\phi_a$. 

In contrast to minimal SU(5), which is broken by an adjoint Higgs representation, FSU(5) is broken to the Standard Model gauge group by a pair of 10-dimensional Higgs representations:
\begin{eqnarray}
H({\bf 10},1)=\left\{Q_H,D^c_H,N_H^c\right\} \; , \quad \quad \bar H(\bar {\bf 10},-1)=\left\{\bar Q_H,\bar D^c_H,\bar N_H^c\right\}~.
\end{eqnarray}
The MSSM Higgs bosons are embedded in another pair of Higgs representations:
\begin{eqnarray}
h({\bf 5},-2)= \left\{T_{H_c},H_d\right\} \; , \quad \quad \bar h(\bar {\bf 5},2)=\left\{ \bar T_{\bar H_c}, H_u\right\} \, ,
\end{eqnarray}
where $T_{H_c}$ and $T_{\bar H_c}$ denote color triplets, 
and $H_d$ and $H_u$ the MSSM Higgs doublets. 

The conventional electroweak hypercharge is a linear combination of the U(1)$_X$ gauge symmetry 
and the diagonal U(1) subgroup of SU(5), namely
\begin{eqnarray}
\frac{Y}{2}= \frac{1}{\sqrt{15}}Y_{24} +\sqrt{\frac{8}{5}} Q_X \, ,
\end{eqnarray}
where the $Q_X$ charge is in units of $\frac{1}{\sqrt{40}}$ and
\begin{eqnarray}
Y_{24}=\sqrt{\frac{3}{5}} {\rm diag}\left(\frac{1}{3},\frac{1}{3},\frac{1}{3},-\frac{1}{2},-\frac{1}{2}\right)~.
\end{eqnarray}
The gauge bosons that get masses
from the breaking of SU(5)$\times$U(1)$\to$ SU(3)$\times$SU(2)$\times$U(1) are $X(3,2)_{1/3},\bar X(\bar 3,2)_{-1/3}$  and a singlet $V_1$, with masses
\begin{eqnarray}
M_X=g_5V \; , \quad \quad M_{V_1}=\sqrt{\frac{5}{2}}\left(\frac{24}{25}g_5^2+\frac{1}{25}g_X^2\right)^{1/2} V \, ,
\label{eq:MX0}
\end{eqnarray}
where the vev $V=\langle N^c_{H_1}\rangle =\langle N^c_{H_2}\rangle$.
The superpotential for this theory is
\begin{align} \notag
W &=  \lambda_1^{ij} F_iF_jh + \lambda_2^{ij} F_i\bar{f}_j\bar{h} +
 \lambda_3^{ij}\bar{f}_i\ell^c_j h + \lambda_4 HHh + \lambda_5
 \bar{H}\bar{H}\bar{h}\\ 
&\quad   + \lambda_6^{ia} F_i\bar{H}\phi_a + \lambda_7^a h\bar{h}\phi_a
 + \lambda_8^{abc}\phi_a\phi_b\phi_c + \mu_\phi^{ab}\phi_a\phi_b\,, 
\label{Wgen} 
\end{align}
where the indices $i,j$ run over the three fermion families, the indices $a, b, c$
have ranges $\ge 3$, and for
simplicity we have suppressed gauge group indices.
We impose
a $\mathbb{Z}_2$ symmetry  
$H\rightarrow -H$
to prevent the mixing of Standard Matter fields with Higgs colour triplets and
elements of the Higgs decuplets. This symmetry also suppresses the
supersymmetric mass term for $H$ and $\bar{H}$, and thus suppresses dimension-five proton decay operators.
The first three terms of the superpotential (\ref{Wgen}) provide the Standard Model Yukawa couplings. The splitting of the triplet and doublet masses in the Higgs 5-plets is accomplished naturally by the
fourth and fifth terms in (\ref{Wgen}),
as these terms yield masses only for the color triplets:
\begin{eqnarray}
M_{H_C}=4\lambda_4V \quad \quad \quad M_{\bar H_C}=4\lambda_5 V \, .
\end{eqnarray}
The sixth term accounts for neutrino masses. The seventh term plays the role of the MSSM
$\mu$-term. The last two terms may play roles in cosmological inflation, along with $\lambda_6$, and 
also play roles in neutrino masses.
GUT symmetry breaking, inflation, leptogenesis, and the generation of neutrino masses in this model have been discussed 
recently in~\cite{EGNNO2,FSU5Cosmo,building}.

The gauge and superpotential couplings of FSU(5) are matched to those of the MSSM at a renormalization scale, $M_{GUT}$, defined to be the scale where $g_2=g_3$ \cite{emo3}:
\bea
 \alpha_2=\alpha_3 = \alpha_5 &,&
 25\alpha_1^{-1} = 24\alpha_X^{-1}+\alpha_5^{-1}  \, , \nonumber \\
 h_t = h_{\nu} = \lambda_2 /\sqrt{2} &,&
 h_b = 4 \lambda_1 \, , \nonumber \\
 h_\tau = \lambda_3 &,& 
\label{matching1}
\eea
where $\alpha_1\equiv (5/3) g_Y^2/(4\pi)$.
Here we quote just the tree-level matching conditions, but our calculations 
include one-loop threshold corrections when the input universality scale, $M_{in}$, is above $M_{GUT}$, which will be discussed separately 
in a more general 
study \cite{eenno2}. 
We note that, unlike minimal SU(5), the neutrino Yukawa couplings are naturally fixed 
to be equal to the up-quark
Yukawa couplings. This is a consequence of the flipping that puts the right-handed neutrinos  
into decuplets in FSU(5), instead of being singlets as
in minimal SU(5), where their Yukawa couplings would be viewed as independent parameters.

The following GUT-scale parameters characterize the FSU(5) GUT model we study.
As mentioned above, we include two independent gaugino masses, a common mass $M_5$ 
for the SU(5) gauginos $\tilde g, \tilde W$ and $\tilde B$, and an independent mass $M_{X1}$
for the `external' gaugino $\tilde B_X$. We also include three independent soft
supersymmetry-breaking scalar masses, $m_{10}$ for sfermions in the $\mathbf{10}$ 
representations of SU(5), $m_5$ for sfermions in the $\mathbf{\overline 5}$
representations of SU(5), and $m_1$ for the right-handed sleptons in the singlet
representations. All of these sfermion mass parameters are assumed to be generation-independent,
and the trilinear soft supersymmetry-breaking parameters $A_0$ are assumed to be universal.
As in the NUHM \cite{nuhm2}, we also assume independent soft supersymmetry-breaking for the $\mathbf{5}$
and $\mathbf{\overline 5}$ Higgs representations, $m_{H_{1,2}}$, and treat the ratio of Higgs vevs,
$\tan \beta$, as a free parameter. Finally, we assume that the Higgs mixing parameter $\mu > 0$,
so as to obtain a supersymmetric contribution to $g_\mu - 2$ with the `interesting'
positive sign.

The matching conditions for the the soft supersymmetry-breaking terms at $M_{GUT}$ are
\bea
 M_2=M_3 = M_{5} &,&
 25M_1\alpha_1^{-1} = 24M_{X1} \alpha_X^{-1}+M_5\alpha_5^{-1} \, , \nonumber \\
 m_{Q}^2=m_{D}^2=m_{N}^2 = m_{10}^2 &,&
m_{U}^2=m_{L}^2 = m_{5}^2 \, , \nonumber \\
 m_{E}^2 = m_{1}^2 &,&  \nonumber \\
 m_{H_u}^2 = m_{h_2}^2 &,&
 m_{H_d}^2 = m_{h_1}^2 \, , \nonumber \\
 A_t=A_\nu = A_b = A_\tau = A_0 &.&  
\label{matching2}
\eea
Once again, these are the tree-level matching conditions, though our calculations include
the one-loop threshold corrections when  $M_{in} > M_{GUT}$ and will be discussed separately in a more general study \cite{eenno2}.
Full universality (as considered in \cite{emo3})
would set $M_5 = M_{X1} = m_{1/2}$ and 
$m_{10} = m_{5} = {m_1} = m_{h_1} = m_{h_2} = m_0$.

Minimization of the Higgs potential determines $\mu$ and the $B$-term at the electroweak scale. This also determines the pseudoscalar Higgs mass, $M_A$,
which we use as an input to {\tt FeynHiggs~2.18.0}~\cite{FH}
to determine the masses of the remaining physical Higgs degrees of freedom.~\footnote{Equivalently,
as in~\cite{nuhm2}, one can treat $\mu$ and $M_A$
as input parameters and use the minimization conditions to solve for the two Higgs soft masses. This approach is taken here as it is more convenient when searching for parameter sets yielding a substantial contribution to 
$g_\mu - 2$.}
Our FSU(5) model is therefore completely specified by the following set of parameters:
\beq
M_5, \, M_{X1}, \, m_{10}, \, m_5, \, m_{1}, \, \mu, \, M_A, \, A_0, \, \tan\beta \, .
\label{CFSU5params}
\eeq
If one were to assume universality at some high input scale, $M_{in} > M_{GUT}$,
additional FSU(5) couplings such as $\lambda_4$, $\lambda_5$ and $\lambda_6$
would also need to be specified, and the relevant RGEs for flipped SU(5) were given in \cite{emo3}. However, here it is assumed that $M_{in} = M_{GUT}$, 
so these parameters are unimportant for the results discussed here, except for the proton lifetime, 
which depends on $\lambda_{4,5}$. We need also to specify the mass of the heaviest left-handed neutrino,
$m_{\nu_3}$, which we take to be 0.05~eV. This and $\lambda_6$ fix the right-handed neutrino mass 
and $\mu_\phi$. However, our results
are quite insensitive to these choices.

Maximization of the supersymmetric contribution to $g_\mu - 2$ requires, {\it a priori},
that either the $\tilde \mu_R$ and the lightest neutralino $\chi$ and/or 
the $\tilde \nu$ and the lighter chargino
must be relatively light, i.e., $\lesssim 1$~TeV. The light $\chi$/$\tilde \mu_R$ option is
favoured in FSU(5) by the fact that the U(1) gaugino mass and $m_1$ are independent of the other soft
supersymmetry-breaking masses and relatively unconstrained, whereas the SU(2) gaugino mass is related by
universality and the standard renormalization calculation to the gluino mass, which is strongly
constrained by fruitless LHC searches~\cite{LHCSUSY}, and the sneutrino mass is likewise
constrained by lower limits on the right-handed up-squark mass. Therefore, we do not pursue the light chargino/$\tilde \nu$ option,
but focus on the light $\chi$/$\tilde \mu_R$ option. Our computation of $\Delta a_\mu$ follows the analysis in \cite{ENO}, which is based on calculations in \cite{IN}.

\section{Results of FSU(5) Parameter Scan}

We report now the results of a scan over the following ranges of the FSU(5) model parameters:
\begin{eqnarray}
M_5 \; \in [1800, 5000]~{\rm GeV}, & M_1 \; \in [100, 1000]~{\rm GeV}\, , \\
M_A \; \in [1500, 3000]~{\rm GeV}, & \mu \; \in [500, 5000]~{\rm GeV} \, , \\ 
m_{10} \; \in [-1000, 4000]~{\rm GeV}, & m_{\overline{5}} \; \in [-500, 1500]~{\rm GeV}\, , \\
m_1 \; \in [-500, 1500]~{\rm GeV}, & A/M_5 \; \in [0, 2] \, , \\
\tan \beta \; \in [35, 40] \, , &
\end{eqnarray}
including $2.2\times10^6$ points~\footnote{Negative values of
soft supersymmetry-breaking
scalar masses should be understood as $m^2/\sqrt{|m^2|}$. Such negative values are consistent
with CMSSM-like phenomenology \cite{feng,mc10}
and with
standard cosmology if the Standard Model vacuum is relatively long-lived when
any charge- and/or colour-breaking minima occur~\cite{flat,EGLOS}.}.

In making our scan, we implement the neutralino LSP requirement $m_{\ell_R} > m_\chi$.
As mentioned above, we assume universality between the values of $m_1$ for the different singlet sleptons,
so we consider the strongest available constraints across the $\ell_R$ of different generations, which are
generally found for the $\tilde e_R$. LEP experiments established lower limits on $m_{\tilde e_R}$ that depend on
other sparticle masses, in particular $m_\chi$~\cite{PDG}. We assume
a LEP lower limit of $100$~GeV in general, reducing to 73~GeV when $m_{\tilde \mu_R} - m_\chi \lesssim 2$~GeV. At the LHC, ATLAS has established the lower limit $m_{\tilde \ell_R} \gtrsim 450$~GeV
when $m_\chi = 0$, where $\ell = e, \mu$,
falling to $\gtrsim 200$~GeV when $m_\chi \simeq 180$~GeV~\cite{slepton}, 
but these lower limits on the $m_{\ell_R}$ are absent for $m_\chi > 180$~GeV. 
An additional LHC constraint is present for compressed spectra when $m_{\mu_R} - m_\chi 
\lesssim 15$~GeV~\cite{Aad:2019qnd}, which is maximized when 
$m_{\mu_R} - m_\chi \simeq 10$~GeV in which  case it excludes $m_{\mu_R} \lesssim 150$~GeV. 
Therefore, in order to maximize the supersymmetric contribution to $g_\mu - 2$ 
we prioritize the region of
parameter space where $m_\chi + 15~{\rm GeV} < m_{\mu_R} \sim 100$~GeV, 
which constrains primarily $m_1$ and $M_{X_1}$.
The other soft supersymmetry-breaking parameters are constrained primarily by 
unsuccessful LHC searches, and we also apply the constraint that $m_h$ calculated 
using {\tt FeynHiggs~2.18.0}~\cite{FH} is within 3~GeV of
the measured Higgs mass. 

We do not use the relic neutralino density as a constraint, since the
flipped SU(5) GUT model contains mechanisms for generating large amounts of entropy~\cite{FSU5Cosmo}. Nevertheless,
in the regions of parameter space that provide the most sizeable contributions to $\Delta a_\mu$, the lightest neutralino (typically mostly a bino) and the right-handed selectron and smuon are quite close in mass, $m_{\mu_R} - m_\chi \simeq 15 - 20$ GeV. In this case,
the neutralino relic density is controlled by slepton coannihilation, which
yields a relic density that is close to that needed to account for the
cold dark matter density determined by 
recent microwave background analyses \cite{Planck} (see also~\cite{Cox:2021gqq}). 

\begin{figure}[!htb]
\begin{center}
\hspace{-.5cm}
\includegraphics[height=69mm]{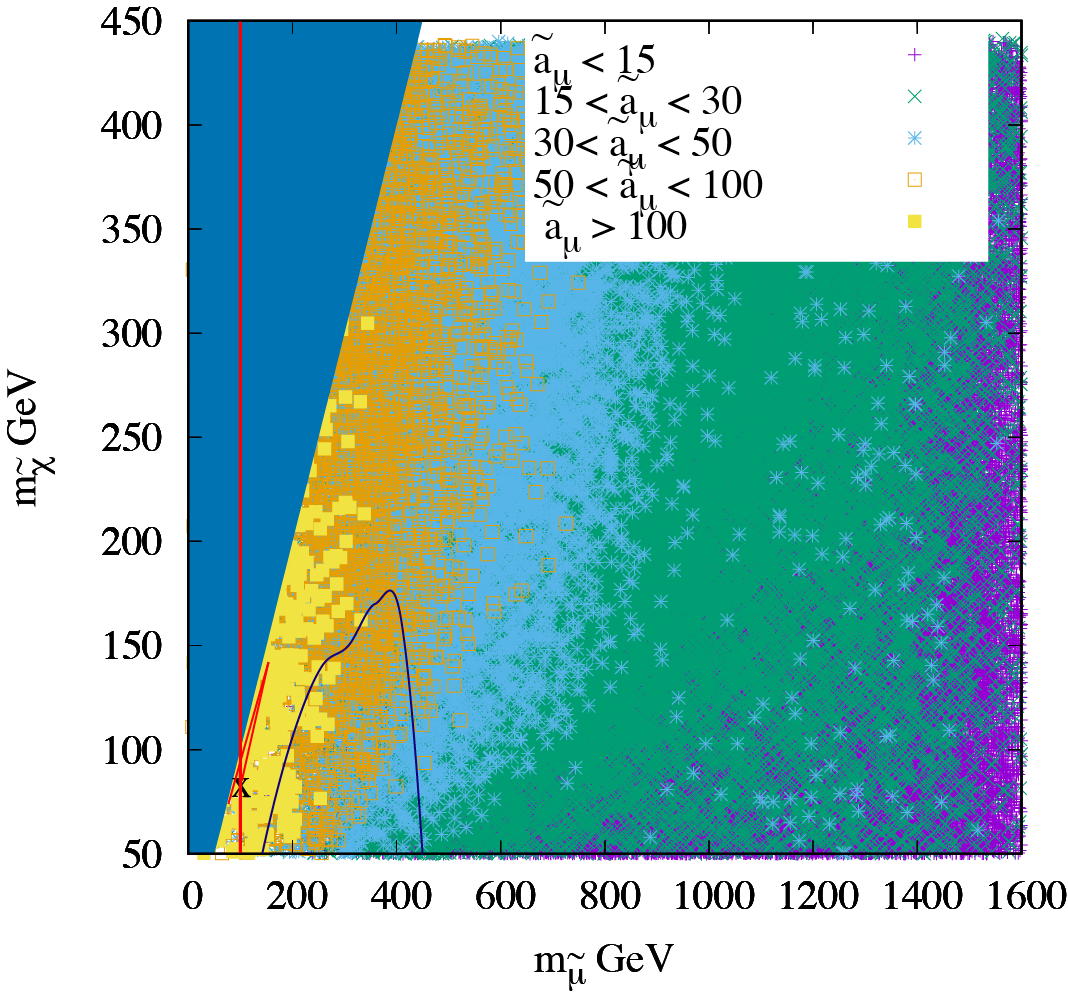}
\hspace{-1cm}
\includegraphics[height=70mm]{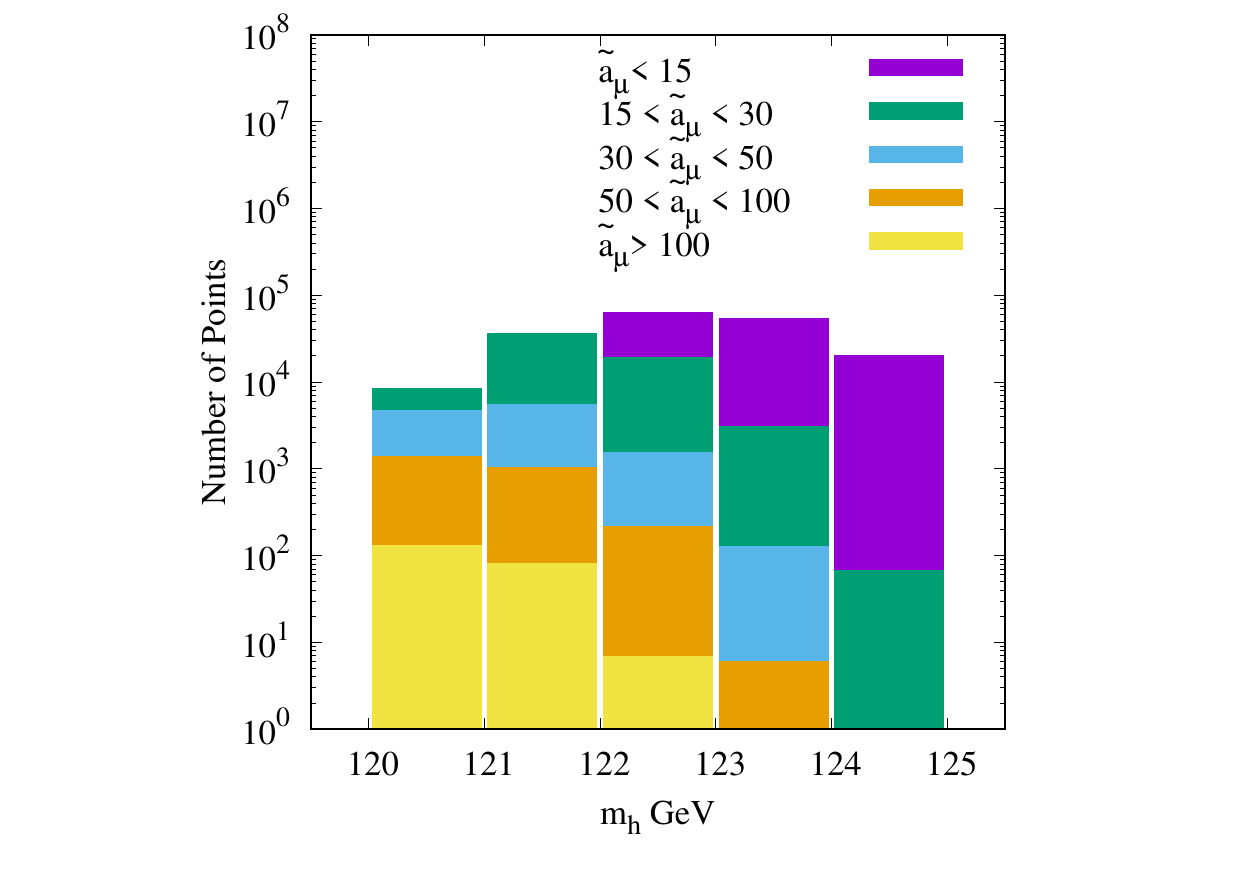}
\caption{\it Left panel: Scatter plot of flipped SU(5) points in the
$(m_{\tilde \mu_R}, m_\chi)$ plane,
color-coded according to the values of the supersymmetric
contribution to $a_\mu$, $\tilde a_\mu \equiv \Delta a_\mu \times 10^{-11}$, that they yield, as
indicated in the legend. The diagonal line represents the
constraint that the LSP is not charged, and the vertical
line represents the LEP lower limit on the slepton mass~\cite{PDG}.
Also visible at small masses are the LHC constraints on
$\tilde \ell_R \to \ell \chi$ where $\ell = e, \mu$~\cite{slepton}. 
The point with the largest value of 
$\Delta a_\mu = 150 \times 10^{-11}$ is indicated with a cross.
Right panel: Stacked histograms of the numbers of points with 
$\tilde a_\mu$ and $m_h$ in the indicated ranges.}
\label{fig:smuchi}
\end{center}
\end{figure}

The left panel of Fig.~\ref{fig:smuchi} shows a scatter plot of FSU(5) 
points in the $(m_{\tilde \mu_R}, m_\chi)$ plane
color-coded according to the values of the supersymmetric contribution to $a_\mu$ that they yield, as
indicated in the legend. 
The darker blue shading covers points with $m_{\tilde \mu_R} < m_\chi$, 
which are therefore excluded because the LSP is charged.
The vertical red line represents the LEP constraint $m_{\tilde e_R} \gtrsim 100$~GeV~\cite{PDG}, where we recall that
$m_{\tilde \mu_R} = m_{\tilde e_R}$ within the approximations we use. 
Also visible at $m_{\tilde \mu_R} \lesssim 450$~GeV is the principal LHC Run~2 constraint
on $\tilde \ell_R \to \ell \chi$ decay~\cite{slepton}, where $\ell = e, \mu$
(blue line), and the additional constraint for $m_{\tilde \mu_R} < 150$~GeV
and small $m_{\tilde \mu_R}- m_\chi$~\cite{Aad:2019qnd} (red line). 
We see that points yielding $\Delta a_\mu > 50 \, (100) \times 10^{-11}$, indicated by orange 
(yellow) boxes, are concentrated at $m_{\tilde \mu_R}, m_\chi \lesssim 500 \, (250)$~GeV.
We note that most of the points with supersymmetric contributions $\Delta a_\mu \gtrsim 100 \times 10^{-11}$
are allowed by the constraints mentioned above. In a dedicated study
we found the largest value
$\Delta a_\mu = 150 \times 10^{-11}$ for the point indicated by a black cross.~\footnote{As mentioned above, 
the limit $m_{\tilde \mu_R} > 100$~GeV is relaxed 
to $m_{\tilde \mu_R} \gtrsim 73$~GeV when $m_{\tilde \mu_R} - m_\chi \lesssim 2$~GeV~\cite{Aad:2019qnd}.
In a dedicated study of this exceptional region we found points with values of
$\Delta a_\mu \gtrsim 220 \times 10^{-11}$.}

The right panel of Fig.~\ref{fig:smuchi} displays stacked histograms of the numbers of points
yielding values of $\tilde a_\mu \equiv \Delta a_\mu \times 10^{11}$ within the indicated ranges,
binned according to the corresponding values of $m_h$ calculated using {\tt FeynHiggs~2.18.0}.
We note that all the points with $\Delta a_\mu > 100 \times 10^{-11}$ correspond to $m_h < 123$~GeV. All points with $m_h > 122$~GeV are allowed if one adopts
a conservative estimate of 3~GeV for the 2-$\sigma$ uncertainty in the
calculation of $m_h$. However, we note that the {\tt FeynHiggs~2.18.0}
code~\cite{FH} returns a 1-$\sigma$ uncertainty in $m_h$ that is
below 1~GeV for the points of greatest interest for $g_\mu - 2$.
We find for scan points with $m_h > 123 \, (124)$~GeV the following maximum values
$\Delta a_\mu = 71 \, (25) \times 10^{-11}$.

\begin{figure}[!htb]
\begin{center}
\hspace{-.5cm}
\includegraphics[height=68mm]{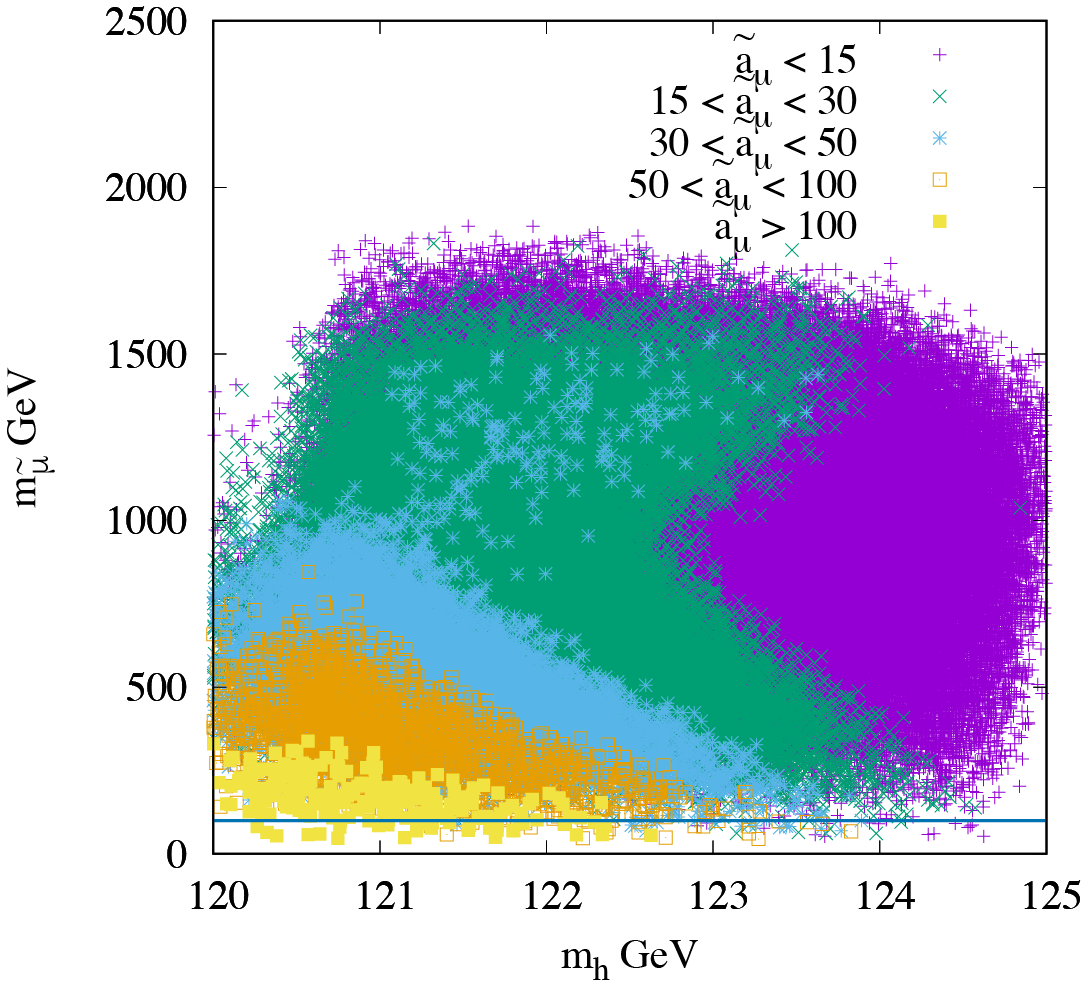}
\includegraphics[height=68mm]{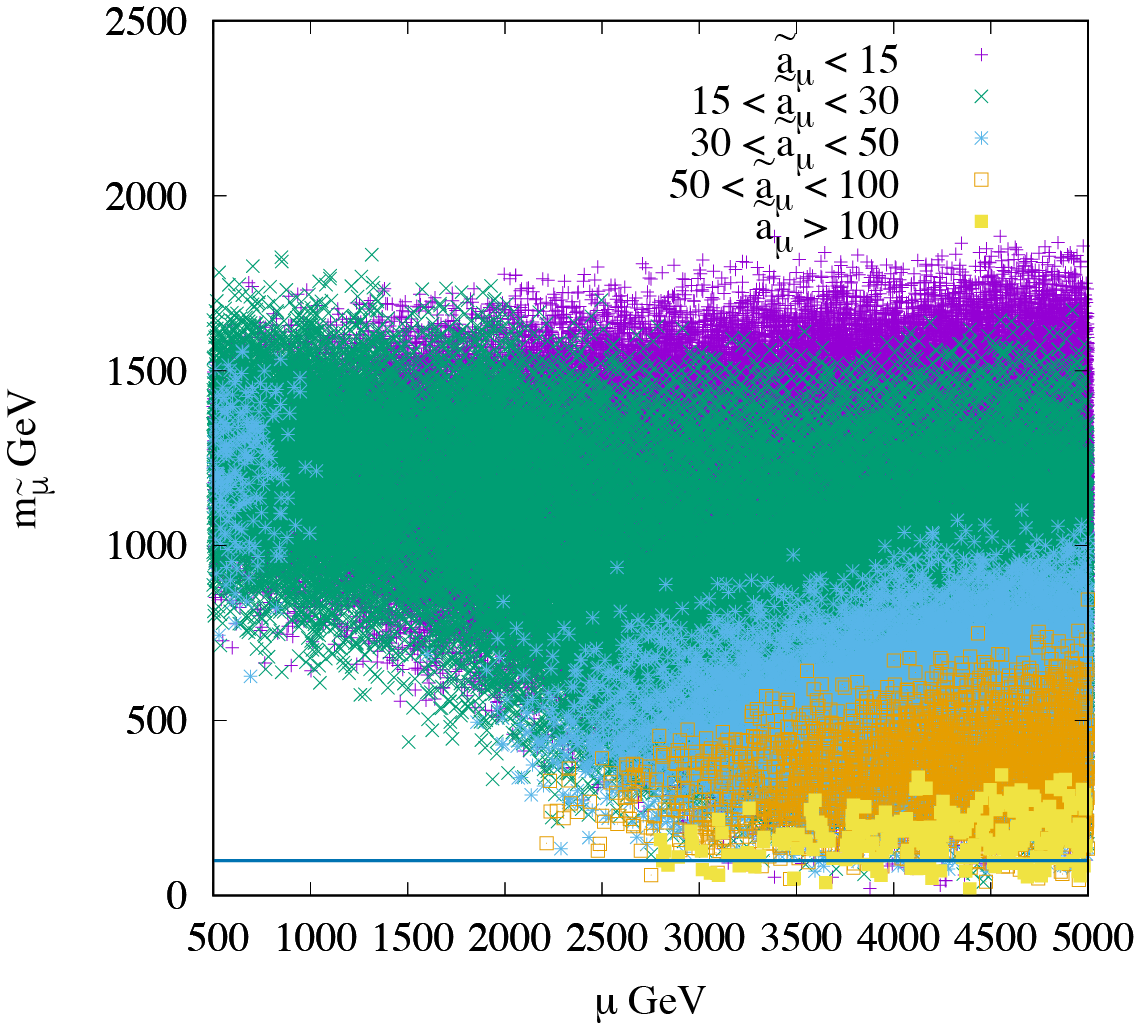}
\caption{\it Scatter plots of flipped SU(5) points in 
(left panel) the $(m_h, m_{\tilde \mu_R})$ plane
and (right panel) the $(\mu, m_{\tilde \mu_R})$ plane,
color-coded according to the values of the supersymmetric
contribution to $a_\mu$ that they yield, as
indicated in the legend. The horizontal
lines represent the LEP lower limit on the slepton mass~\cite{PDG}.}
\label{fig:smuchi2}
\end{center}
\end{figure}

The left panel of Fig.~\ref{fig:smuchi2} shows a scatter
plot of FSU(5) points in the $(m_h, m_{\tilde \mu_R})$
plane and color-coded as in Fig.~\ref{fig:smuchi}.
The horizontal line represents the LEP lower limit on the slepton
mass of 100~GeV~\cite{PDG}. We see that the the values of
$\Delta a_\mu$ tend to decrease with increasing $m_{\tilde \mu_R}$
and $m_h$. The trend with $m_{\tilde \mu_R}$ was seen already in the
left panel of Fig.~\ref{fig:smuchi}, and the trend with $m_h$ reflects
the fact that larger values of $m_h$ correspond in general to larger
sparticle masses, in particular $\tilde \mu_L$. This suppresses
$\tilde \mu_L/\tilde \mu_R$ mixing and hence the 
corresponding contribution to $\Delta a_\mu$.
The right panel of Fig.~\ref{fig:smuchi2} shows a scatter
plot of flipped SU(5) points in the $(\mu, m_{\tilde \mu_R})$
plane, where we see that the points yielding $\Delta a_\mu 
\gtrsim 50 \times 10^{-11}$ correspond to relatively large
values of $\mu > 2500$~GeV, where the
$\tilde \mu_R/\tilde \mu_L$ mixing
contribution is enhanced.

\begin{figure}[htb]
\begin{center}
\includegraphics[height=80mm]{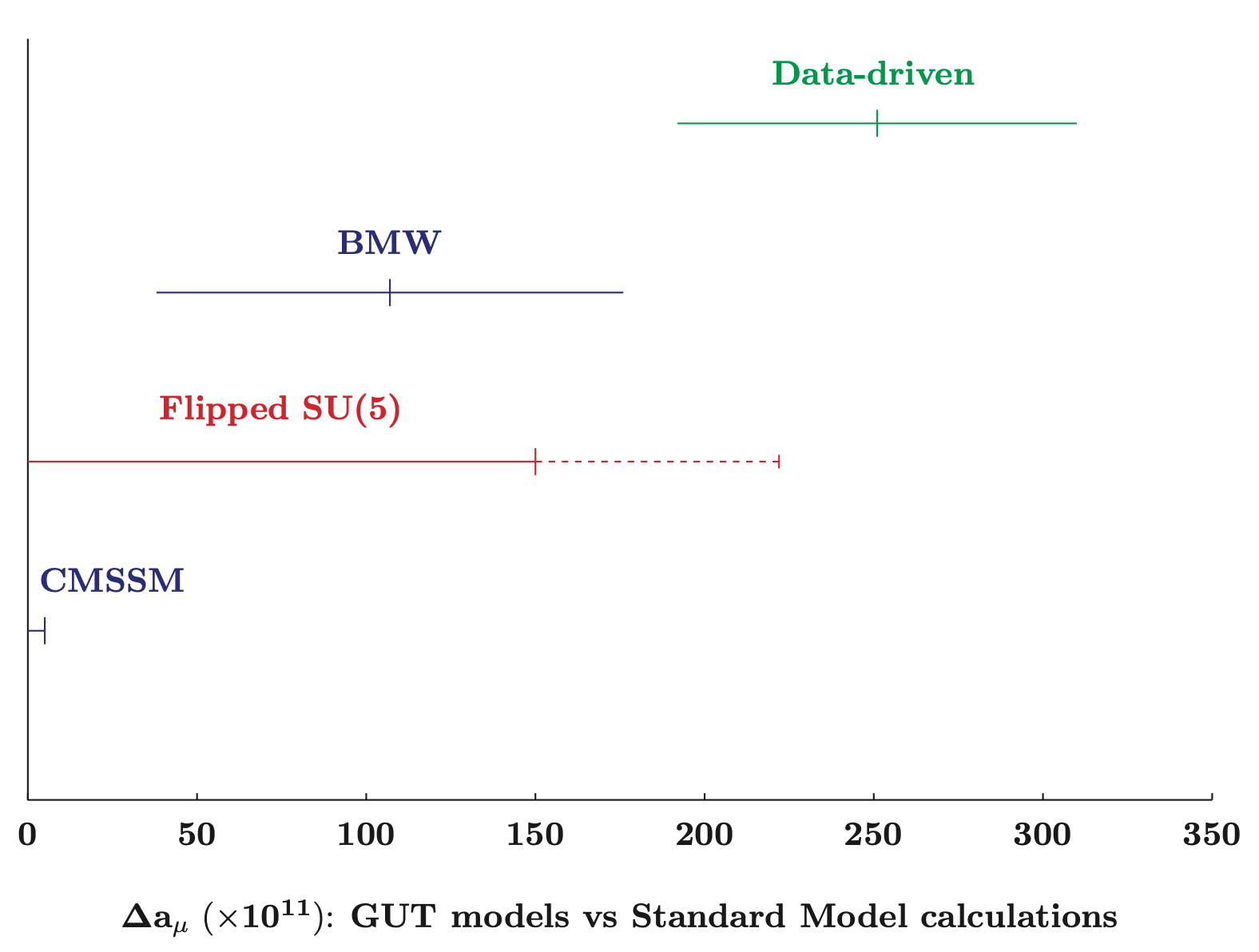}
\caption{\it Comparison of the ranges of the discrepancy in $a_\mu$
between the combination of the BNL and Fermilab measurements with the
data-driven estimate taken from the Theory Initiative~\cite{Theory} 
(green line),
from the BMW lattice calculation~\cite{Lattice} (black range), 
and the ranges found
in flipped SU(5) in this paper (red range, general region shown as solid line,
extension in exceptional region shown dashed) and in the CMSSM~\cite{CMSSM} 
(blue range).}
\label{fig:Comparison}
\end{center}
\end{figure}

Fig.~\ref{fig:Comparison} compares the ranges of the discrepancy, $\Delta a_\mu$
between the combination of the BNL and Fermilab measurements and the
data-driven estimate of $a_\mu$ taken from the Theory Initiative~\cite{Theory} 
(green line) and the BMW lattice calculation~\cite{Lattice} (black line), 
together with the range of the supersymmetric contribution to $\Delta a_\mu$
found in our general scan of the flipped SU(5) parameter space (red line). We see that the flipped
SU(5) model could resolve completely the residual 1.5-$\sigma$ discrepancy
between the BMW lattice calculation~\cite{Lattice} and the experimental measurements.
It also reduces
the discrepancy between the data-driven Standard Model estimate and the measurements
to less than 2 standard deviations.~\footnote{The red dashed line shows the additional
range of $\Delta a_\mu$ that is found in the exceptional region where
$m_{\tilde \mu_R} - m_\chi \lesssim 2$~GeV and $m_{\tilde \mu_R} \gtrsim 73$~GeV.}
Also shown is the 2-$\sigma$ range of $\Delta a_\mu$
found in a 
global analysis of the CMSSM that includes all relevant
constraints from LHC Run 2, previous experiments and
constraints on dark matter~\cite{CMSSM} (blue line).
We see that the supersymmetric contribution to
$\Delta a_\mu$ in the CMSSM is $\sim 30$ times
smaller than in flipped SU(5), and is negligible compared to the
experimental discrepancies with the Standard Model calculations.

As has been mentioned above, the generic FSU(5) point that makes the largest contribution to
$a_\mu$ yields $\Delta a_\mu = 150 \times 10^{-11}$. Table~\ref{tab:numbers} shows
the input parameters for this point, including those pertaining to the specification of
the GUT model~\footnote{The GUT mass scales are largely determined by extrapolation from low-energy data, and are insensitive to the values of $\lambda_{4,5,6}$. Our results are also insensitive to
$m_{\nu_3}$ within the range allowed by cosmological data.} 
and those pertaining to the supersymmetry scales. We also list in
Table~\ref{tab:numbers} the output MSSM particle masses and other observables. 
We observe that, apart from the lightest neutralino LSP and the $\tilde \mu_R$ 
(and the near-degenerate $\tilde e_R$),~\footnote{Note that the $\tilde{\tau}_R$ is 
much heavier than the $\tilde \mu_R$ and $\tilde e_R$, because $m_{H_d}^2$ has large 
negative values, which increase $m_{\tilde{\tau}_R}$ at low energies.}
the squarks and gluinos are in general far
beyond the current reach of the LHC~\cite{LHCSUSY} and even the prospective reach of the HL-LHC~\cite{HL-LHC},
though within reach of FCC-hh~\cite{FCC-hh} or SppC~\cite{SppC}. This
is a general feature of points that yield interesting values of $\Delta a_\mu$ and
$m_h > 122$~GeV. The optimal point is also compatible with the LHC Run~2 limits in the 
$(M_{A}, \tan \beta)$ plane \cite{ATLAS:2020zms}. The $\tilde \mu_R$ and $\tilde e_R$ might be 
within reach of future LHC searches via conventional missing-energy signatures~\cite{slepton}
and/or dedicated searches in the compressed spectrum region~\cite{Aad:2019qnd},
possibly using the LHC as a photon collider~\cite{Beresford:2018pbt}.
They could also be within the reach of
an $e^+e^-$ collider operating at 250~GeV in the center of mass, such as the ILC~\cite{ILC},
FCC-ee~\cite{FCC-ee} or CEPC~\cite{CEPC}.

\begin{table}
\small
\begin{center}
\label{tbl:MassesFV}
\begin{tabular}{|ccc|}
\hline 

\multicolumn{3}{|c|}{Input GUT parameters (masses in units of $10^{16}$ GeV)} \\
\hline
$M_{GUT} = 1.00$  & $M_X = 0.79$ & $V= 1.13$ \\
$\lambda_4 = 0.1$  & $\lambda_5 = 0.3$ &  $\lambda_6 = 0.001$ \\
$g_5 = 0.70$  & $g_X = 0.70$  & $m_{\nu_3} = 0.05$~eV \\
\hline
\multicolumn{3}{|c|}{Input supersymmetry parameters (masses in GeV units)}\\
\hline
$M_5 = 2460$ & $M_1 = 240$ & $\mu = 4770$\\
 $m_{10} = 930$ & $m_{\overline{5}} =450$ & $m_1 = 0$ \\
$M_A = 2100$ &  $A_0/M_5 = 0.67$ & $\tan \beta = 35$ \\
\hline
\multicolumn{3}{|c|}{MSSM particle masses (in GeV units)} \\
\hline
$m_\chi =84$ &   $m_{\tilde t_1} = 4030 $ & $m_{\tilde g} = 5090 $  \\
$m_{\chi_2} = 2160$ & $m_{\chi_3} = 5080$ & $m_{\chi_4} = 5080$  \\
$m_{\tilde \mu_R} = 101$ & $m_{\tilde \mu_L} = 1600$ & $m_{\tilde \tau_1} = 1010$ \\
$m_{\tilde q_L} = 4470$ & $m_{\tilde d_R} = 4250$ & $m_{\tilde u_R} = 4170$ \\
$m_{\tilde t_2} = 4410 $ & $m_{\tilde b_1} = 4170$ & $m_{\tilde b_2} = 4400 $ \\
$m_{\chi^\pm} = 2160$ & $m_{H,A} = 2100$ & $m_{H^\pm} = 2100$ \\
\hline
\multicolumn{3}{|c|}{Other observables} \\
\hline
$\Delta a_\mu = 150 \times 10^{-11}$ & $\Omega_\chi h^2 = 0.13$ & $m_h = 122$ GeV \\
Normal-ordered $\nu$ masses: & $\tau_{p \to e^+ \pi^0}|_{_{\rm NO}} = 4.6 \times 10^{35}$~yrs &
$\tau_{p \to \mu^+ \pi^0}|_{_{\rm NO}} = 4.7 \times 10^{36}$~yrs \\
Inverse-ordered $\nu$ masses: & $\tau_{p \to e^+ \pi^0}|_{_{\rm IO}} = 1.4 \times 10^{37}$~yrs &
$\tau_{p \to \mu^+ \pi^0}|_{_{\rm IO}} = 9.8 \times 10^{35}$~yrs \\
\hline
\end{tabular}
\end{center}
\caption{\it Parameters and predictions of an FSU(5)
point that yields $\Delta a_\mu = 150 \times 10^{-11}$.
\label{tab:numbers}}
\end{table}

Finally, we also show in Table~\ref{tab:numbers} the values of some other
observables for this point. The relic LSP density $\Omega_\chi h^2$ calculated
assuming adiabatic cosmological evolution happens to fall quite close to
the range of cold dark matter density favoured by Planck~\cite{Planck} and other 
measurements, though this was not imposed {\it a priori}. 
This is because smaller values of 
$\Omega_\chi h^2$ are allowed if there is another source of cold dark matter,
while a complete FSU(5) model of cosmology favours a large amount of entropy
generation that would dilute even a quite substantial potential overdensity
of LSPs \cite{FSU5Cosmo}. However, for the point whose parameters are given in Table~\ref{tab:numbers}
and other, similar points, LSP coannihilations with the $\tilde \mu_R$ and $\tilde e_R$
naturally bring $\Omega_\chi h^2$ close to or within the range preferred by Planck even before any
such entropy generation \cite{Cox:2021gqq}. We also show in Table~\ref{tab:numbers} predictions for
the partial lifetimes for $p \to e^+ \pi^0$ and $p \to \mu^+ \pi^0$ in variants
of the FSU(5) in which the light neutrino masses are ordered either normally (NO) 
or inversely (IO). We see that in all cases these partial lifetimes are well beyond the present experimental
limits and the prospective reach of planned experiments such as Hyper-Kamiokande.

\section{Summary}

We have explored in this paper the range of possible values of
the supersymmetric contribution to $a_\mu$ in the flipped SU(5)
GUT model. This model has more parameters than the familiar CMSSM, or
even a standard SU(5) GUT. Specifically, there are two
independent gaugino mass parameters in flipped SU(5), one for
the SU(5) adjoint gauginos, $M_{5}$, and another, $M_{X1}$, for the gaugino
corresponding to the external U(1) factor. This decouples the
mass of the lightest neutralino LSP from those of the gluino and
the SU(2) gauginos. Also, flipped SU(5) has three independent soft
supersymmetry-breaking scalar masses per generation, for the $\mathbf{10}, \mathbf{\overline{5}}$
and singlet representations of SU(5), $m_{10}, m_{\overline{5}}$ and $m_1$,
compared to two parameters in standard SU(5) or just
one mass parameter in the CMSSM. Moreover, since the supersymmetric partner of the right-handed
muon is in a singlet representation of flipped SU(5), its mass avoids the constraints imposed
on squarks and left-handed sleptons. The freedom in the choice of $M_{X1}$ and $m_1$ allows
the LSP and the $\tilde \mu_R$ to be much lighter than the other sparticles, opening up the
possibility of a much larger contribution to $a_\mu$ than in the CMSSM, for example.

Indeed, we have found that the flipped SU(5) contribution to $a_\mu$ could be as
large as $\sim 150 \times 10^{-11}$, even after taking the available LEP and LHC constraints
into account, whereas these constraints favour values $\lesssim 5 \times 10^{-11}$ in the CMSSM.
The potential flipped SU(5) contribution to $a_\mu$ would reduce the discrepancy between experiment
and the data-driven calculation of the Standard Model contribution to below 2 standard deviations,
and be completely consistent with the central value of the BMW lattice calculation.
Flipped SU(5) is therefore an example of a GUT-based supersymmetric model that may bridge the gap
between experiment and the Standard Model.

We have also discussed in this paper some other possible experimental signatures of this flipped
SU(5) scenario for $g_\mu - 2$. The lightest supersymmetric particles, namely the lightest
neutralino, the $\tilde e_R$ and the $\tilde \mu_R$ may all be detectable in
dedicated searches at the LHC, or in experiments at a 250-GeV $e^+ e^-$
collider such as the ILC, FCC-ee or CEPC. On the other hand, the heavier supersymmetric particles
would be beyond the reach of the LHC, and their detection would have to wait for FCC-hh or SppC.
Suitable neutrino masses can be incorporated, with either normal or inverse mass ordering.
In both cases, the flipped SU(5) model predicts a proton lifetime well beyond the current
constraints and also beyond the reach of planned experiments. We note also that the
cross section for spin-dependent dark matter scattering is far below the current experimental
limit.

We will return soon to these and other issues in a more detailed study of the phenomenology of flipped SU(5)~\cite{eenno2}. 

\section*{Acknowledgements}
The work of J.E. was supported partly by the United Kingdom STFC Grant
ST/T000759/1 and partly by the Estonian Research Council via a Mobilitas
Pluss grant. The work of N.N. was supported by the Grant-in-Aid
for Young Scientists (No.21K13916), Innovative Areas
(No.18H05542), and Scientific Research B (No.20H01897). The work of D.V.N. was supported partly by the DOE grant
DE-FG02-13ER42020 and partly by the Alexander S. Onassis Public Benefit
Foundation. The work of K.A.O. was supported partly by the DOE grant
DE-SC0011842 at the University of Minnesota.



\begin{thebibliography}{99}

\bibitem{BNL1}
H.~N.~Brown \textit{et al.} [Muon g-2 Collaboration],
Phys. Rev. Lett. \textbf{86} (2001), 2227-2231
[arXiv:hep-ex/0102017 [hep-ex]].

\bibitem{BNL2}
G.~W.~Bennett \textit{et al.} [Muon g-2 Collaboration],
Phys. Rev. D \textbf{73} (2006), 072003
[arXiv:hep-ex/0602035 [hep-ex]].

\bibitem{FNAL}
B. Abi \textit{et al.} [Muon g-2 Collaboration]
Phys.~Rev.~Lett. \textbf{126} (2021), 141801
[arXiv:2104.03281 [hep-ex]].

\bibitem{Theory}
T.~Aoyama, \textit{et al.}
Phys. Rept. \textbf{887} (2020), 1-166
[arXiv:2006.04822 [hep-ph]].

\bibitem{ENO}
J.~R.~Ellis, D.~V.~Nanopoulos and K.~A.~Olive,
Phys. Lett. B \textbf{508} (2001), 65-73
[arXiv:hep-ph/0102331 [hep-ph]].

\bibitem{g-2}
L.~L.~Everett, G.~L.~Kane, S.~Rigolin and L.~Wang,
Phys.\ Rev.\ Lett.\  {\bf 86} (2001) 3484 
[arXiv:hep-ph/0102145];
J.~L.~Feng and K.~T.~Matchev,
Phys.\ Rev.\ Lett.\  {\bf 86} (2001) 3480 
[arXiv:hep-ph/0102146];
E.~A.~Baltz and P.~Gondolo,   
Phys.\ Rev.\ Lett.\  {\bf 86} (2001) 5004 
[arXiv:hep-ph/0102147];
U.~Chattopadhyay and P.~Nath,
Phys.\ Rev.\ Lett.\  {\bf 86} (2001) 5854 
[arXiv:hep-ph/0102157];
S.~Komine, T.~Moroi and M.~Yamaguchi,
Phys.\ Lett.\ B {\bf 506} (2001) 93 
[arXiv:hep-ph/0102204];
J.~Hisano and K.~Tobe,
Phys. Lett. B \textbf{510}, 197-204 (2001)
[arXiv:hep-ph/0102315 [hep-ph]];
R.~Arnowitt, B.~Dutta, B.~Hu and Y.~Santoso,
Phys.\ Lett.\ B {\bf 505} (2001) 177  
[arXiv:hep-ph/0102344]
S.~P.~Martin and J.~D.~Wells,
Phys.\ Rev.\ D {\bf 64} (2001) 035003 
[arXiv:hep-ph/0103067];
H.~Baer, C.~Balazs, J.~Ferrandis and X.~Tata,
Phys.\ Rev.\ D {\bf 64} (2001) 035004 
[arXiv:hep-ph/0103280].


\bibitem{LHCSUSY}
For a compendium of ATLAS searches for supersymmetry, see\\
{\tt https://twiki.cern.ch/twiki/bin/view/AtlasPublic/SupersymmetryPublicResults};
for a compendium of CMS searches for supersymmetry, see\\
{\tt https://twiki.cern.ch/twiki/bin/view/CMSPublic/PhysicsResultsSUS}.

\bibitem{cmssm0}
  M.~Drees and M.~M.~Nojiri,
Phys.\ Rev.\ D {\bf 47} (1993) 376 [arXiv:hep-ph/9207234];
 G.~L.~Kane, C.~F.~Kolda, L.~Roszkowski and J.~D.~Wells,
  Phys.\ Rev.\  D {\bf 49} (1994) 6173
  [arXiv:hep-ph/9312272];
J.~R.~Ellis, K.~A.~Olive, Y.~Santoso and V.~C.~Spanos,
Phys.\ Lett.\ B {\bf 565} (2003) 176
[arXiv:hep-ph/0303043];
H.~Baer and C.~Balazs,
  JCAP {\bf 0305}, 006 (2003)
  [arXiv:hep-ph/0303114];
  A.~B.~Lahanas and D.~V.~Nanopoulos,
  Phys.\ Lett.\  B {\bf 568}, 55 (2003)
  [arXiv:hep-ph/0303130];
U.~Chattopadhyay, A.~Corsetti and P.~Nath,
  Phys.\ Rev.\  D {\bf 68}, 035005 (2003)
  [arXiv:hep-ph/0303201];
     J.~Ellis and K.~A.~Olive,
  arXiv:1001.3651 [astro-ph.CO], published in {\it Particle dark matter}, ed. G.~Bertone, pp. 142-163;
  J.~Ellis and K.~A.~Olive,
  Eur.\ Phys.\ J.\ C {\bf 72}, 2005 (2012)
  [arXiv:1202.3262 [hep-ph]];
  J.~Ellis, F.~Luo, K.~A.~Olive and P.~Sandick,
Eur. Phys. J. C \textbf{73}, no.4, 2403 (2013)
[arXiv:1212.4476 [hep-ph]];
O.~Buchmueller {\it et al.},
  Eur.\ Phys.\ J.\ C {\bf 74} (2014) 3,  2809
  [arXiv:1312.5233 [hep-ph]];
O.~Buchmueller, M.~Citron, J.~Ellis, S.~Guha, J.~Marrouche, K.~A.~Olive, K.~de Vries and J.~Zheng,
Eur. Phys. J. C \textbf{75}, no.10, 469 (2015)
[erratum: Eur. Phys. J. C \textbf{76}, no.4, 190 (2016)]
[arXiv:1505.04702 [hep-ph]].
E.~A.~Bagnaschi \textit{et al.}
Eur. Phys. J. C \textbf{75}, 500 (2015)
[arXiv:1508.01173 [hep-ph]];
J.~Ellis, J.~L.~Evans, F.~Luo, N.~Nagata, K.~A.~Olive and P.~Sandick,
Eur. Phys. J. C \textbf{76}, no.1, 8 (2016)
[arXiv:1509.08838 [hep-ph]];
J.~Ellis, J.~L.~Evans, F.~Luo, K.~A.~Olive and J.~Zheng,
Eur. Phys. J. C \textbf{78}, no.5, 425 (2018)
[arXiv:1801.09855 [hep-ph]];
  E.~Bagnaschi, H.~Bahl, J.~Ellis, J.~Evans, T.~Hahn, S.~Heinemeyer, W.~Hollik, K.~Olive, S.~Passehr, H.~Rzehak, I.~Sobolev, G.~Weiglein and J.~Zheng,
Eur. Phys. J. C \textbf{79}, no.2, 149 (2019)
[arXiv:1810.10905 [hep-ph]];
J.~Ellis, J.~L.~Evans, N.~Nagata, K.~A.~Olive and L.~Velasco-Sevilla,
Eur. Phys. J. C \textbf{80}, no.4, 332 (2020)
[arXiv:1912.04888 [hep-ph]].

\bibitem{CMSSM}
E.~Bagnaschi \textit{et al.}
Eur. Phys. J. C \textbf{77} (2017) no.2, 104
[arXiv:1610.10084 [hep-ph]].
  
\bibitem{otherCMSSM}
See also 
P.~Athron, C.~Bal\'azs, D.~H.~Jacob, W.~Kotlarski, D.~St\"ockinger and H.~St\"ockinger-Kim,
[arXiv:2104.03691 [hep-ph]];
F.~Wang, L.~Wu, Y.~Xiao, J.~M.~Yang and Y.~Zhang,
arXiv:2104.03262 [hep-ph];
M.~Chakraborti, L.~Roszkowski and S.~Trojanowski,
JHEP \textbf{05} (2021), 252
[arXiv:2104.04458 [hep-ph]].

  
\bibitem{pMSSM11}
E.~Bagnaschi {\it et al.},
  Eur.\ Phys.\ J.\ C {\bf 78} (2018) no.3,  256
  [arXiv:1710.11091 [hep-ph]].
  
\bibitem{Sven}
See also M.~Chakraborti, S.~Heinemeyer and I.~Saha,
arXiv:2104.03287 [hep-ph];
arXiv:2105.06408 [hep-ph].

\bibitem{otherSUSY}
M.~Endo, K.~Hamaguchi, S.~Iwamoto and T.~Kitahara,
arXiv:2104.03217 [hep-ph];
S.~Iwamoto, T.~T.~Yanagida and N.~Yokozaki,
[arXiv:2104.03223 [hep-ph]];
Y.~Gu, N.~Liu, L.~Su and D.~Wang,
arXiv:2104.03239 [hep-ph];
W.~Yin,
JHEP \textbf{06} (2021), 029
[arXiv:2104.03259 [hep-ph]];
M.~Abdughani, Y.~Z.~Fan, L.~Feng, Y.~L.~Sming Tsai, L.~Wu and Q.~Yuan,
arXiv:2104.03274 [hep-ph];
M.~Ibe, S.~Kobayashi, Y.~Nakayama and S.~Shirai,
arXiv:2104.03289 [hep-ph];
S.~Heinemeyer, E.~Kpatcha, I.~Lara, D.~E.~L\'opez-Fogliani, C.~Mu\~noz and N.~Nagata,
[arXiv:2104.03294 [hep-ph]];
S.~Baum, M.~Carena, N.~R.~Shah and C.~E.~M.~Wagner,
arXiv:2104.03302 [hep-ph];
H.~B.~Zhang, C.~X.~Liu, J.~L.~Yang and T.~F.~Feng,
arXiv:2104.03489 [hep-ph];
W.~Ahmed, I.~Khan, J.~Li, T.~Li, S.~Raza and W.~Zhang,
arXiv:2104.03491 [hep-ph];
A.~Aboubrahim, M.~Klasen and P.~Nath,
arXiv:2104.03839 [hep-ph];
H.~Baer, V.~Barger and H.~Serce,
arXiv:2104.07597 [hep-ph];
W.~Altmannshofer, S.~A.~Gadam, S.~Gori and N.~Hamer,
arXiv:2104.08293 [hep-ph];
A.~Aboubrahim, P.~Nath and R.~M.~Syed,
JHEP \textbf{06}, 002 (2021)
[arXiv:2104.10114 [hep-ph]];
K.~S.~Jeong, J.~Kawamura and C.~B.~Park,
arXiv:2106.04238 [hep-ph];
Z.~Li, G.~L.~Liu, F.~Wang, J.~M.~Yang and Y.~Zhang,
arXiv:2106.04466 [hep-ph].


\bibitem{Cox:2021gqq}
P.~Cox, C.~Han and T.~T.~Yanagida,
arXiv:2104.03290 [hep-ph].

\bibitem{FSU5}
S.~M.~Barr,
  Phys.\ Lett.\  {\bf 112B} (1982) 219;
  S.~M.~Barr,
  Phys.\ Rev.\ D {\bf 40}, 2457 (1989);
J.~P.~Derendinger, J.~E.~Kim and D.~V.~Nanopoulos,
  Phys.\ Lett.\  {\bf 139B} (1984) 170;
I.~Antoniadis, J.~R.~Ellis, J.~S.~Hagelin and D.~V.~Nanopoulos,
  Phys.\ Lett.\ B {\bf 194} (1987) 231;
  Phys.\ Lett.\ B {\bf 205} (1988) 459;
  Phys.\ Lett.\ B {\bf 208} (1988) 209
   Addendum: [Phys.\ Lett.\ B {\bf 213} (1988) 562];
  Phys.\ Lett.\ B {\bf 231} (1989) 65.


\bibitem{LNW}
J.~L.~Lopez, D.~V.~Nanopoulos and X.~Wang,
Phys. Rev. D \textbf{49} (1994), 366-372
[arXiv:hep-ph/9308336 [hep-ph]].

\bibitem{Lattice}
 S.~Borsanyi, 
 \textit{et al.}
Nature \textbf{593}, no.7857, 51-55 (2021)
[arXiv:2002.12347 [hep-lat]].

\bibitem{pMSSM}
See, for example,
C.~F.~Berger, J.~S.~Gainer, J.~L.~Hewett and T.~G.~Rizzo,
  JHEP {\bf 0902}, 023 (2009)
  [arXiv:0812.0980 [hep-ph]];
S.~S.~AbdusSalam, B.~C.~Allanach, F.~Quevedo, F.~Feroz and M.~Hobson,
  Phys.\ Rev.\ D {\bf 81}, 095012 (2010)
  [arXiv:0904.2548 [hep-ph]];
  J.~A.~Conley, J.~S.~Gainer, J.~L.~Hewett, M.~P.~Le and T.~G.~Rizzo,
  Eur.\ Phys.\ J.\ C {\bf 71}, 1697 (2011)
  [arXiv:1009.2539 [hep-ph]];
  J.~A.~Conley, J.~S.~Gainer, J.~L.~Hewett, M.~P.~Le and T.~G.~Rizzo,
  [arXiv:1103.1697 [hep-ph]];
  B.~C.~Allanach, A.~J.~Barr, A.~Dafinca and C.~Gwenlan,
  JHEP {\bf 1107}, 104 (2011)
  [arXiv:1105.1024 [hep-ph]];
  S.~Sekmen, S.~Kraml, J.~Lykken, F.~Moortgat, S.~Padhi, L.~Pape, M.~Pierini and H.~B.~Prosper {\it et al.},
  JHEP {\bf 1202} (2012) 075
  [arXiv:1109.5119 [hep-ph]];
  A.~Arbey, M.~Battaglia and F.~Mahmoudi,
  Eur.\ Phys.\ J.\ C {\bf 72} (2012) 1847
  [arXiv:1110.3726 [hep-ph]];
  A.~Arbey, M.~Battaglia, A.~Djouadi and F.~Mahmoudi,
  Phys.\ Lett.\ B {\bf 720} (2013) 153
  [arXiv:1211.4004 [hep-ph]];
  M.~W.~Cahill-Rowley, J.~L.~Hewett, A.~Ismail and T.~G.~Rizzo,
  Phys.\ Rev.\ D {\bf 88} (2013) 3,  035002
  [arXiv:1211.1981 [hep-ph]];
C.~Strege, G.~Bertone, G.~J.~Besjes, S.~Caron, R.~Ruiz de Austri, A.~Strubig and R.~Trotta,
  JHEP {\bf 1409} (2014) 081
  [arXiv:1405.0622 [hep-ph]];
  M.~Cahill-Rowley, J.~L.~Hewett, A.~Ismail and T.~G.~Rizzo,
  Phys.\ Rev.\ D {\bf 91} (2015) 5,  055002
  [arXiv:1407.4130 [hep-ph]];
  L.~Roszkowski, E.~M.~Sessolo and A.~J.~Williams,
  JHEP {\bf 1502}, 014 (2015)
  [arXiv:1411.5214 [hep-ph]];
M.~E.~Cabrera-Catalan, S.~Ando, C.~Weniger and F.~Zandanel,
Phys. Rev. D \textbf{92}, no.3, 035018 (2015)
[arXiv:1503.00599 [hep-ph]];
J.~Chakrabortty, A.~Choudhury and S.~Mondal,
JHEP \textbf{07}, 038 (2015)
[arXiv:1503.08703 [hep-ph]].
  
  \bibitem{mc11}
K.~J.~de Vries {\it et al.},
  Eur.\ Phys.\ J.\ C {\bf 75} (2015) no.9,  422
  [arXiv:1504.03260 [hep-ph]].

\bibitem{nuhm2}
J.~Ellis, K.~Olive and Y.~Santoso,
Phys.\ Lett.\  B~{\bf 539} (2002) 107
[arXiv:hep-ph/0204192];
J.~R.~Ellis, T.~Falk, K.~A.~Olive and Y.~Santoso,
Nucl.\ Phys.\ B {\bf 652} (2003) 259
[arXiv:hep-ph/0210205].

\bibitem{IN}
T.~Ibrahim and P.~Nath,
Phys.\ Rev.\ {\bf D62} (2000) 015004.

  \bibitem{EGNNO2}
  J.~Ellis, M.~A.~G.~Garc{\' i}a, N.~Nagata, D.~V.~Nanopoulos and K.~A.~Olive,
  JCAP {\bf 1707} (2017) no.07,  006
  [arXiv:1704.07331 [hep-ph]];
  J.~Ellis, M.~A.~Garc{\' i}a, N.~Nagata, D.~V.~Nanopoulos and K.~A.~Olive,
JCAP \textbf{04}, 009 (2019)
[arXiv:1812.08184 [hep-ph]];
  J.~Ellis, M.~A.~Garc{\' i}a, N.~Nagata, D.~V.~Nanopoulos and K.~A.~Olive,
Phys. Lett. B \textbf{797}, 134864 (2019)
[arXiv:1906.08483 [hep-ph]].


\bibitem{FSU5Cosmo}
J.~Ellis, M.~A.~G.~Garcia, N.~Nagata, D.~V.~Nanopoulos and K.~A.~Olive,
JCAP \textbf{01} (2020), 035
[arXiv:1910.11755 [hep-ph]].

\bibitem{building}
J.~Ellis, M.~A.~G.~Garcia, N.~Nagata, D.~V.~Nanopoulos, K.~A.~Olive and S.~Verner,
Int. J. Mod. Phys. D \textbf{29}, no.16, 2030011 (2020)
[arXiv:2009.01709 [hep-ph]].

\bibitem{emo3} 
  J.~Ellis, A.~Mustafayev and K.~A.~Olive,
  Eur.\ Phys.\ J.\ C {\bf 71}, 1689 (2011)
  [arXiv:1103.5140 [hep-ph]].
  
  \bibitem{eenno2}
  J.~Ellis, J.~L.~Evans, N.~Nagata, D.~V.~Nanopoulos, and K.~A.~Olive, (in preparation). 


  
\bibitem{FH}
  S.~Heinemeyer, W.~Hollik and G.~Weiglein,
  Comput.\ Phys.\ Commun.\  {\bf 124} (2000) 76
  [arXiv:hep-ph/9812320];
  S.~Heinemeyer, W.~Hollik and G.~Weiglein,
  Eur.\ Phys.\ J.\ C {\bf 9} (1999) 343
  [arXiv:hep-ph/9812472];
  G.~Degrassi, S.~Heinemeyer, W.~Hollik, P.~Slavich and G.~Weiglein,
  Eur.\ Phys.\ J.\ C {\bf 28} (2003) 133
  [arXiv:hep-ph/0212020];
  M.~Frank {\it et al.},
  JHEP {\bf 0702} (2007) 047
  [arXiv:hep-ph/0611326];
  T.~Hahn, S.~Heinemeyer, W.~Hollik, H.~Rzehak and G.~Weiglein,
  Comput.\ Phys.\ Commun.\  {\bf 180} (2009) 1426;
  T.~Hahn, S.~Heinemeyer, W.~Hollik, H.~Rzehak and G.~Weiglein,
  Phys.\ Rev.\ Lett.\  {\bf 112} (2014) 14,  141801
  [arXiv:1312.4937 [hep-ph]];
  H.~Bahl and W.~Hollik,
  Eur.\ Phys.\ J.\ C {\bf 76} (2016) no.9,  499
  [arXiv:1608.01880 [hep-ph]];
  H.~Bahl, S.~Heinemeyer, W.~Hollik and G.~Weiglein,
  Eur.\ Phys.\ J.\ C {\bf 78} (2018) no.1,  57
  [arXiv:1706.00346 [hep-ph]].
  H.~Bahl, T.~Hahn, S.~Heinemeyer, W.~Hollik, S.~Pa\ss ehr, H.~Rzehak and G.~Weiglein,
  Comput.\ Phys.\ Commun.\  {\bf 249} (2020) 107099
  [arXiv:1811.09073 [hep-ph]].
  See {\tt http://www.feynhiggs.de} for updates.

 \bibitem{feng}
   J.~L.~Feng, A.~Rajaraman and B.~T.~Smith,
  Phys.\ Rev.\ D {\bf 74}, 015013 (2006)
  [hep-ph/0512172];
  A.~Rajaraman and B.~T.~Smith,
  Phys.\ Rev.\ D {\bf 75}, 115015 (2007)
  [hep-ph/0612235].
  
  \bibitem{mc10}
 O.~Buchmueller {\it et al.},
  Eur.\ Phys.\ J.\ C {\bf 74}, no. 12, 3212 (2014)
  [arXiv:1408.4060 [hep-ph]].
  
  \bibitem{flat}
T.~Falk, K.~A.~Olive, L.~Roszkowski and M.~Srednicki,
  Phys.\ Lett.\ B {\bf 367}, 183 (1996)
  [hep-ph/9510308];
 T.~Falk, K.~A.~Olive, L.~Roszkowski, A.~Singh and M.~Srednicki,
  Phys.\ Lett.\ B {\bf 396}, 50 (1997)
  [hep-ph/9611325].


\bibitem{EGLOS}
J.~R.~Ellis, J.~Giedt, O.~Lebedev, K.~Olive and M.~Srednicki,
Phys. Rev. D \textbf{78} (2008), 075006
[arXiv:0806.3648 [hep-ph]].

\bibitem{PDG}
P.~A.~Zyla \textit{et al.} [Particle Data Group],
PTEP \textbf{2020} (2020) no.8, 083C01.

\bibitem{slepton}
G.~Aad \textit{et al.} [ATLAS Collaboration],
Eur. Phys. J. C \textbf{80} (2020) no.2, 123
[arXiv:1908.08215 [hep-ex]].
  
\bibitem{Aad:2019qnd}
G.~Aad \textit{et al.} [ATLAS],
Phys. Rev. D \textbf{101}, no.5, 052005 (2020)
[arXiv:1911.12606 [hep-ex]].
  
  \bibitem{Planck}
  N.~Aghanim \textit{et al.} [Planck Collaboration],
Astron. Astrophys. \textbf{641}, A6 (2020)
[arXiv:1807.06209 [astro-ph.CO]].



\bibitem{HL-LHC}
  X.~Cid Vidal, \textit{et al.}
{\it Report from Working Group 3: Beyond the Standard Model physics at the HL-LHC and HE-LHC},
CERN Yellow Rep. Monogr. \textbf{7} (2019), 585-865
[arXiv:1812.07831 [hep-ph]].

  \bibitem{FCC-hh}
  A.~Abada \textit{et al.} [FCC Collaboration],
{\it FCC-hh: The Hadron Collider: Future Circular Collider Conceptual Design Report Volume 3},
Eur. Phys. J. ST \textbf{228} (2019) no.4, 755-1107.

\bibitem{SppC}
J.~Tang \textit{et al.}
{\it Concept for a Future Super Proton-Proton Collider},
arXiv:1507.03224 [physics.acc-ph].

\bibitem{ATLAS:2020zms}
G.~Aad \textit{et al.} [ATLAS],
Phys. Rev. Lett. \textbf{125}, no.5, 051801 (2020)
[arXiv:2002.12223 [hep-ex]].

\bibitem{Beresford:2018pbt}
L.~Beresford and J.~Liu,
Phys. Rev. Lett. \textbf{123}, no.14, 141801 (2019)
arXiv:1811.06465 [hep-ph].

\bibitem{ILC}
H.~Baer \textit{et al.}
{\it The International Linear Collider Technical Design Report - Volume 2: Physics},
arXiv:1306.6352 [hep-ph].

\bibitem{FCC-ee}
A.~Abada \textit{et al.} [FCC Collaboration],
{\it FCC-ee: The Lepton Collider: Future Circular Collider Conceptual Design Report Volume 2},
Eur. Phys. J. ST \textbf{228} (2019) no.2, 261-623.

\bibitem{CEPC}
J.~B.~Guimar\~aes da Costa \textit{et al.} [CEPC Study Group],
{\it CEPC Conceptual Design Report: Volume 2 - Physics \& Detector},
arXiv:1811.10545 [hep-ex].

\end{thebibliography}
\end{document}